\documentclass[conference]{IEEEtran}

\usepackage{color}
\usepackage{amsmath,amssymb,graphicx,amsthm}
\def\MR#1{}
\usepackage[hidelinks]{hyperref}

\theoremstyle{definition}

\def\MR#1{}

\begin{document}

\title{Image recovery from rotational and \\ translational invariants}

\author{
	\IEEEauthorblockN{Nicholas F. Marshall\IEEEauthorrefmark{1}, Ti-Yen Lan\IEEEauthorrefmark{2}, Tamir Bendory\IEEEauthorrefmark{3}, Amit Singer\IEEEauthorrefmark{1}\IEEEauthorrefmark{2}}
	\IEEEauthorblockA{\IEEEauthorrefmark{1}Department of Mathematics, Princeton University, Princeton, NJ, USA}
	\IEEEauthorblockA{\IEEEauthorrefmark{2}The Program in Applied and Computational Mathematics, Princeton University, Princeton, NJ, USA}
	\IEEEauthorblockA{\IEEEauthorrefmark{3}School of Electrical Engineering, Tel Aviv University, Tel Aviv, Israel}
}

\maketitle


\begin{abstract}
We introduce a framework for recovering an image from its rotationally 
and translationally invariant features based on autocorrelation analysis. 
This work is an instance of
the multi-target detection statistical model, which is
mainly used to study the mathematical and computational properties of
single-particle reconstruction using cryo-electron microscopy (cryo-EM)
at low signal-to-noise ratios. We demonstrate with synthetic 
numerical experiments that an image can be reconstructed from
rotationally and translationally invariant features and show that the
reconstruction is robust to noise.  These results constitute an important step
towards the goal of structure determination of small biomolecules using cryo-EM.
\end{abstract}


\section{Introduction}
\label{sec:intro}
\subsection{Problem statement} \label{probstate}
We consider the problem of estimating a target image from a large $m \times m$
noisy measurement $M$ that contains $p$ randomly rotated copies of the
target. More precisely, suppose that $f : \mathbb{R}^2 \rightarrow \mathbb{R}$
is a target image supported on the open unit disc, and $f_\phi$ is the rotation
of $f$ by angle $\phi$ about the origin. Let $F_\phi : \mathbb{Z}^2
\rightarrow \mathbb{R}$ be a discrete version of $f_\phi$ defined by
\begin{equation}
F_\phi(\vec{x}) := f_\phi \left(\vec{x}/n \right), \quad \text{for } \vec{x} \in
\mathbb{Z}^2,
\end{equation}
where $n \ll m$ is a fixed positive integer.  We assume the measurement $M :
\{1,\ldots,m\}^2 \rightarrow \mathbb{R}$ is of the form
\begin{equation} \label{eq_M}
M(\vec{x}) := \sum_{j=1}^{p} F_{\phi_j}(\vec{x}-\vec{x}_j) +
\varepsilon(\vec{x}),  \quad \text{for } \vec{x} \in \{1,\ldots,m\}^2,
\end{equation}
where $\phi_1,\ldots,\phi_p \in [0,2\pi)$ are uniformly random angles,
$\vec{x}_1,\ldots,\vec{x}_p \in \{n,\ldots,m-n+1\}^2$ are translations of 
the target images, and $\varepsilon(\vec{x})$ is i.i.d. Gaussian noise with mean $0$
and variance $\sigma^2$. For simplicity, we assume that images in the
measurement are separated by at least one image diameter, see
\eqref{sepcond}. Recent work on a related problem suggests
that this separation restriction can be alleviated \cite{lan2019multi}.

\begin{figure}[t!]
\includegraphics[width=0.4\textwidth]{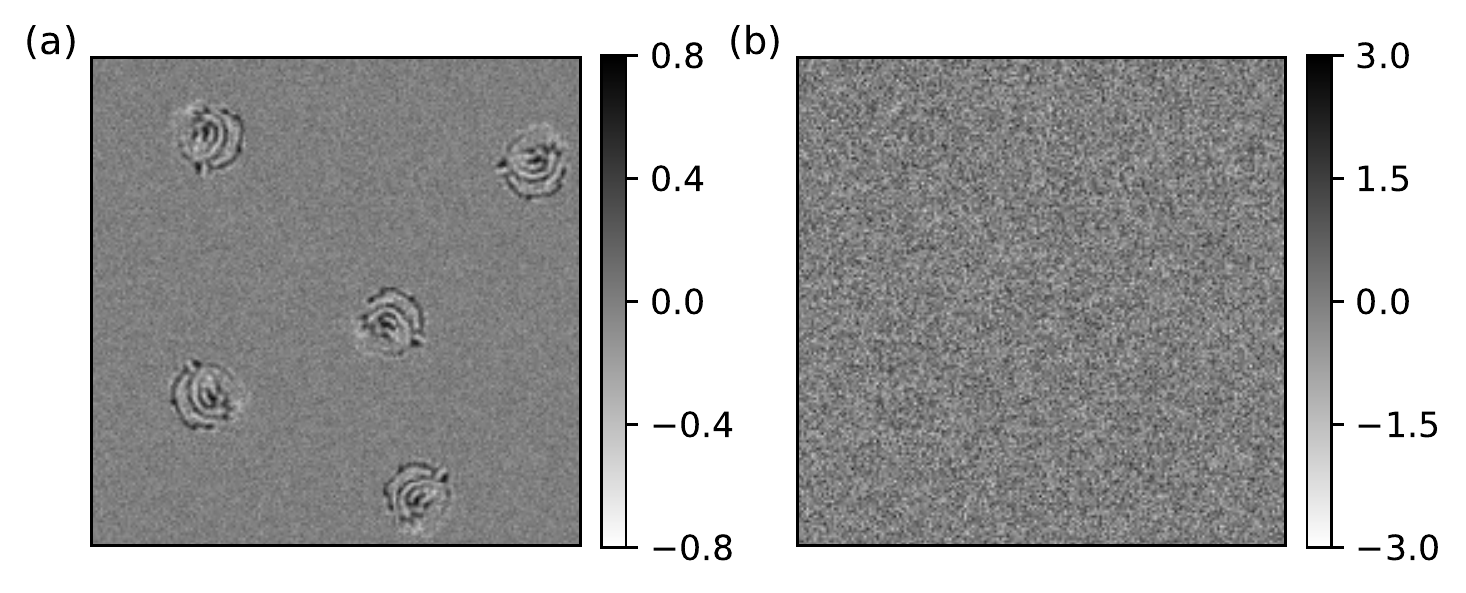}
\centering\caption{Two measurements $M$  with the same five rotated copies of a
target image, but different noise levels: SNR~=~10
(left), and SNR = 0.1 (right).} \label{fig:micrograph}
\end{figure}

Our goal is to recover the image $f$ from the noisy measurement $M$. We
emphasize that with respect to this goal, the rotations $\phi_1,\ldots,\phi_p$
and translations $\vec{x}_1,\ldots,\vec{x}_p$ are nuisance parameters that do
not necessarily need to be estimated. If the signal-to-noise ratio (SNR) is high
as in Figure~\ref{fig:micrograph}(a), then estimating  these rotations and
translations is straightforward, and the image can be recovered by aligning and
averaging its different copies in the measurement. However, if the SNR is
low as in Figure~\ref{fig:micrograph}(b), which is the case of interest in this
study, then this approach may be problematic as even detecting the image
occurrences becomes challenging~\cite{aguerrebere2016fundamental,
bendory2019multi}. 

We are interested in the following question: given a large enough measurement
$M$, can we recover the target image $f$ regardless of the level of noise? More
precisely, if the number of target image occurrences $p$ grows at a constant
rate $\gamma$ with the size $m^2$ of the measurement $M$, can we recover the
image $f$ for any fixed noise variance $\sigma^2$ as $m \rightarrow \infty$? 

\subsection{Motivation}
We are motivated by challenges in single-particle reconstruction using
cryo-electron microscopy (cryo-EM). The measurements in cryo-EM consist of 2-D
tomographic projections of identical biomolecules of unknown 3-D orientations
embedded in a large noisy image, called a micrograph. In the current analysis
workflow of cryo-EM data~\cite{Bendory2019single, frank2006three,
grant2018cistem, punjani2017cryosparc, scheres2012relion, tang2007eman2}, the
2-D projections are first detected and extracted from the micrograph, and later
rotationally and translationally aligned to reconstruct the 3-D molecular
structure.  This approach is problematic for small molecules, which are
difficult to detect due to their lower SNR. This difficulty of detection in 
turn sets a lower bound on the usable molecule size in the current analysis workflow of
cryo-EM data~\cite{henderson1995potential}. To circumvent this fundamental barrier,
recent papers~\cite{bendory2018toward,bendory2019multi} suggest to directly
estimate the 3-D structure from the micrograph, without an intermediate
detection stage; this approach was inspired by Kam~\cite{kam1980reconstruction}
who introduced autocorrelation analysis to cryo-EM.

The estimation problem described in Section \ref{probstate} is a
simplified version of the cryo-EM reconstruction problem: the tomographic
projection operator is omitted and we observe the same \mbox{2-D} image
multiple times with random in-plane rotations. This image recovery problem
is an instance of the \emph{multi-target detection} statistical model, 
in which a set of signals appear multiple times at unknown locations in a noisy 
measurement~\cite{bendory2018toward, bendory2019multi, lan2019multi}.
Here, we extend previous works by taking in-plane rotations into account, which
forms an important step towards the analysis of the full cryo-EM problem.

\subsection{Main contribution}
The principal contribution of this paper is to demonstrate that a target image
$f$ can be recovered from a measurement $M$ of the form \eqref{eq_M} without
estimating the rotations and translations of the images in the measurement. As
a consequence, limits on estimating these nuisance parameters do not impose
limits on estimating the target image $f$. In particular, we empirically
demonstrate that the estimation of $f$ is feasible at any noise level given a
large enough measurement.  From a computational perspective, this approach is
highly efficient as it requires only one pass over the measurement to calculate
the autocorrelations; this is in contrast to the likelihood-based techniques
that may become prohibitive as the data size increases. For a more detailed
discussion, see~\cite{lan2019multi}.  

\section{Rotationally and translationally invariant features}
\label{invariant}
We begin by studying invariance in the continuous setting. The simplest
invariant to both rotations and translations is
$$
s_1 := \int_{\mathbb{R}^2} f(\vec{x}) d\vec{x},
$$
which is the mean of the image; however, clearly more information is needed to
recover the image. Motivated by autocorrelation analysis we next consider the
rotationally-averaged second-order autocorrelation $s_2 : \mathbb{R}^2
\rightarrow \mathbb{R}$ by
$$
s_2(\vec{x}_1) :=  \int_0^{2\pi} \int_{\mathbb{R}^2}
f_\phi(\vec{x}) f_\phi(\vec{x} + \vec{x}_1) d \vec{x}  d \phi.
$$
By a change of variables of integration,  it is clear that $s_2$ is only a
function of $|\vec{x}_1|$, that is, it is a 1-D function, and thus does not
contain enough information to encode a 2-D image. Thus, we proceed to consider
the rotationally-averaged third-order autocorrelation $s_3: \mathbb{R}^2 \times
\mathbb{R}^2 \rightarrow \mathbb{R}$ defined by
\begin{equation} \label{eq:continuous}
s_3(\vec{x}_1,\vec{x}_2) := \int_0^{2 \pi} \int_{\mathbb{R}^2}
f_\phi(\vec{x}) f_\phi(\vec{x}+\vec{x}_1) f_\phi(\vec{x}+\vec{x}_2) d\vec{x} d\phi. \nonumber
\end{equation}
It is straightforward to verify that $s_3$ is a function of three parameters: 
$|\vec{x}_1|$, $|\vec{x}_2|$ and $\theta(\vec{x}_1,\vec{x}_2)$, where
$\theta(\vec{x}_1,\vec{x}_2)$
denotes the angle between the vectors $\vec{x}_1,\vec{x}_2 \in \mathbb{R}^2$.
In this work, we show empirically that $f$ can be recovered from 
$S_3 : \mathbb{Z}^2 \times \mathbb{Z}^2 \rightarrow \mathbb{R}$, which is a
discrete version of $s_3$ defined by
\begin{equation} \label{s3}
S_3(\vec{x}_1,\vec{x}_2) := \int_0^{2\pi} \sum_{\vec{x} \in \mathbb{Z}^2}
F_\phi(\vec{x}) F_\phi(\vec{x} + \vec{x}_1) F_{\phi}(\vec{x}+\vec{x}_2) d\phi.
\end{equation}
We claim that the function $S_3$ can be approximated from the measurement $M$ by
computing the third-order autocorrelation $A_3 : \mathbb{Z}^2 \times
\mathbb{Z}^2 \rightarrow \mathbb{R}$ defined by
\begin{equation} \label{eq:A3}
A_3(\vec{x}_1, \vec{x}_2) := \frac{1}{m^2} \sum_{\vec{x} \in \mathbb{Z}^2} 
M(\vec{x}) M(\vec{x}+\vec{x}_1) M(\vec{x}+\vec{x}_2).
\end{equation}
Indeed, recall that the images are assumed to be separated in the measurement
by one image diameter:
\begin{equation} \label{sepcond}
| \vec{x}_{j_1} - \vec{x}_{j_2} | > 4n, \quad \text{for} \quad j_1 \not =
j_2,
\end{equation}
and that the rotations in the measurement are chosen uniformly at random.
Under these assumptions it is straightforward to show, see for
example~\cite{bendory2018toward}, that if $p/m^2 \rightarrow \gamma$, then 
\begin{multline} \label{A2S}
A_3(\vec{x}_1, \vec{x}_2) \rightarrow 
\frac{\gamma}{2\pi} S_3(\vec{x}_1, \vec{x}_2) \\ + \sigma^2 \frac{\gamma S_1}{2\pi} 
\big( \delta(\vec{x_1}) + \delta(\vec{x_2}) + \delta(\vec{x_1} - \vec{x_2})
\big),
\end{multline}
as $m \rightarrow \infty$ for any fixed noise level $\sigma^2$, where
$$
S_1 := \int_0^{2\pi} \sum_{\vec{x} \in \mathbb{Z}^2} F_\phi(\vec{x}) d\phi,
$$
and $\delta(\vec{x}) = 1$ if $\vec{x} = \vec{0}$ and is zero otherwise. 
Crucially, equation~\eqref{A2S} relates $A_3$, a quantity that can be estimated
accurately from the data, with functions of the target: $S_1$ and $S_3$, while
averaging out the effects of the nuisance variables. 
In practice, $\sigma^2$ and $\gamma S_1$ can be estimated from $M$. More specifically, 
$\sigma^2$ can be estimated by the variance of the pixel values of $M$ in the low SNR 
regime, while $\gamma S_1$ can be estimated by the empirical mean of $M$. As a result, 
$S_3$ can be estimated from $A_3$ up to a constant factor $\gamma$. 

The analysis above motivates the following question: can the image $f$ be
robustly reconstructed from $S_3$? We demonstrate empirically that the answer to
this question is positive.  For simplicity of exposition, we describe a method
of recovering $f$ that just involves $S_3$, although we note that $S_1$ and
$S_2$, which represent the discrete analogs of $s_1$ and $s_2$, may be used to
aid in the reconstruction process. 

The problem of estimating $f$ from $M$, as described in Section \ref{probstate},
should not be confused with the related problem of recovering an image from a
set of measurements $\{M_j\}_{j=1}^n$, each of which has exactly one shifted and
rotated observation. Several moment-based techniques have been proposed to
address this problem at high noise levels, see for instance~\cite{Kondor,
Mallat, ozyesil2018synchronization, sadler1992shift}, but none can be applied
directly to our problem where only a single, large observation $M$ is available. 

\section{Image recovery from invariants}
\subsection{Steerable basis}
Recall that the target image $f : \mathbb{R}^2 \rightarrow \mathbb{R}$ is
supported on the unit disc. We assume that $f$ is bandlimited in the basis of
Dirichlet Laplacian eigenfunctions on the unit disk. In polar coordinates
$(r,\theta)$, these eigenfunctions are of the form 
\begin{equation} \label{eq1}
\psi_{\nu,q}(r,\theta) := J_\nu\left( \lambda_{\nu,q} r \right) e^{i \nu \theta},
\quad \text{for }  (\nu,q) \in \mathbb{Z} \times \mathbb{Z}_{>0},
\end{equation}
where $J_\nu$ is the $\nu$-th order Bessel function of the first kind, and
$\lambda_{\nu,q} > 0$ is the $q$-th positive root of $J_\nu$. The eigenvalue
associated with $\psi_{\nu,q}$ is $\lambda_{\nu,q}^2$, and thus, the assumption
that $f$ is bandlimited can be written as
\begin{equation}
\label{model}
f(r,\theta) =	\sum_{(\nu,q): \lambda_{\nu,q} \le \lambda} \alpha_{\nu,q}
\psi_{\nu,q}(r,\theta), \quad \text{for } r \le 1,
\end{equation}
where $\lambda > 0$ is the bandlimit frequency, and $\alpha_{\nu,q}$ are the
associated expansion coefficients.  If we further define 
$$
g_\nu(r,\theta) = \sum_{q : \lambda_{\nu,q} \le \lambda} 
	\alpha_{\nu,q} \psi_{\nu,q}(r,\theta),
$$
then we can write
\begin{equation} \label{sumg}
f(r,\theta) = \sum_{\nu = -\nu_{\mathrm{max}}}^{\nu_{\mathrm{max}}}
g_\nu(r,\theta), 
\end{equation}
where $\nu_\text{max} := \max \{ \nu : \lambda_{\nu,1} \le \lambda\}$. An
advantage of expressing a function in Dirichlet Laplacian eigenfunctions is
that the basis is steerable in the sense that it diagonalizes the rotation
operator: rotating $f$ by $\phi$ corresponds to multiplying each term
of $g_\nu$ by $e^{i \nu \phi}$:
\begin{equation} \label{grot}
f(r,\theta + \phi) = \sum_{\nu = -\nu_{\mathrm{max}}}^{\nu_{\mathrm{max}}}
g_\nu(r,\theta) e^{i \nu \phi}.
\end{equation}

\subsection{Computing invariants}
Recall that $F_\phi : \mathbb{Z}^2 \rightarrow \mathbb{R}$ is the discretely
sampled version of $f_\phi$, which is defined by
$F_\phi(\vec{x}) = f_\phi \left( \vec{x}/n \right)$ for $\vec{x} \in
\mathbb{Z}^2$. In the following, we consider $F_\phi$ as a function on
$$
\mathcal{J} := \{-2n,\ldots,2n - 1\}^2 \subset \mathbb{Z}^2.
$$ 
Since $f_\phi$ is supported on the open unit disc, it follows that 
$$
\text{supp}(F_\phi) \subset \{ \vec{x} \in \mathcal{J}: |\vec{x}| < n \}. 
$$
Let $\hat{F}_\phi : \mathcal{J} \rightarrow \mathbb{C}$ by
$$
\hat{F}_\phi(\vec{k}) = \sum_{\vec{x} \in \mathcal{J}}
F_\phi(\vec{x}) e^{-2\pi i \vec{k}\cdot\vec{x} / 4n},
\quad \text{for } \vec{k} \in \mathcal{J},
$$
denote the discrete Fourier transform (DFT) of $F_\phi$. Similarly, we can
consider $S_3$ as a function on $\mathcal{J} \times \mathcal{J}$ and define
its DFT $\hat{S}_3 : \mathcal{J} \times \mathcal{J} \rightarrow \mathbb{C}$
by
\begin{equation} \label{dcts}
\hat{S}_3(\vec{k}_1,\vec{k}_2) := \sum_{\vec{x}_1, \vec{x}_2 \in \mathcal{J}}
S_3(\vec{x}_1, \vec{x}_2) e^{-2\pi i (\vec{k}_1\cdot\vec{x}_1 +
\vec{k}_2\cdot\vec{x}_2) / 4n},
\end{equation}
for $\vec{k}_1,\vec{k}_2 \in \mathcal{J}$. By substituting \eqref{s3} into 
\eqref{dcts}, it is straightforward to show that
\begin{equation} \label{fs3}
\hat{S}_3(\vec{k}_1,\vec{k}_2)
= \int_0^{2\pi} \hat{F}_\phi(\vec{k}_1) \hat{F}_\phi(\vec{k}_2)
\hat{F}_\phi(-\vec{k}_1-\vec{k}_2) d\phi. 
\end{equation}
The triple product in \eqref{fs3} corresponds to the Fourier transform 
of the third-order autocorrelation. This triple product is called the bispectrum~\cite{tukey1953spectral}
and many of its analytical and computational properties have been studied; see
for instance~\cite{bendory2017bispectrum,sadler1992shift}. We note that the
bispectrum can be generalized to more involved operations, such as 2-D 
rotations~\cite{ma2018heterogeneous}, 3-D rotations~\cite{Kondor}, 
and general compact groups~\cite{kakarala2012bispectrum}.

Define $\Psi_{\nu, q}: \mathcal{J} \rightarrow \mathbb{C}$ as the discrete
samples of the Dirichlet Laplacian eigenfunctions:
$$
\Psi_{\nu, q}(\vec{x}) = \psi_{\nu,q}(\vec{x}/n), \quad \text{for } \vec{x} \in
\mathcal{J},
$$
where $\psi_{\nu,q}$ is considered as a function supported on the unit disc. 
With this notation, we can express $F_\phi$ as 
\begin{align*}
F_\phi(\vec{x}) &= \sum_{(\nu,q): \lambda_{\nu,q} \le \lambda} \alpha_{\nu,q}
	\Psi_{\nu,q}(\vec{x}) e^{i \nu \phi} \nonumber \\
	&= \sum_{\nu = -\nu_{\mathrm{max}}}^{\nu_{\mathrm{max}}}
	\left( \sum_{q: \lambda_{\nu,q} \le \lambda} \alpha_{\nu,q}
	\Psi_{\nu,q}(\vec{x}) \right) e^{i \nu \phi}.
\end{align*}
As a consequence, the products $F_\phi(\vec{x}) F_\phi(\vec{x} + \vec{x}_1)
F_\phi(\vec{x} + \vec{x}_2)$ that appear in \eqref{s3} are bandlimited by
$3\nu_\mathrm{max}$ with respect to $\phi$. Therefore, we can replace the
integral over $\phi$ in \eqref{fs3} by a summation over angles sampled at the
Nyquist rate, that is,
$$
\hat{S}_3(\vec{k}_1,\vec{k}_2) = \sum_{\nu=0}^{6\nu_\text{max}-1}
\hat{F}_{\phi_\nu}(\vec{k}_1) \hat{F}_{\phi_\nu}(\vec{k}_2)
\hat{F}_{\phi_\nu}(-\vec{k}_1-\vec{k}_2),
$$
where $\phi_\nu := 2\pi \nu/(6 \nu_\mathrm{max})$. By linearity of the DFT, 
we have 
$$
\hat{F}_\phi(\vec{k}) =	\sum_{(\nu,q): \lambda_{\nu,q} \le \lambda}
\alpha_{\nu,q} \hat{\Psi}_{\nu, q}(\vec{k}) e^{i \nu \phi},
$$
where $\hat{\Psi}_{\nu, q}: \mathcal{J} \rightarrow \mathbb{C}$ denotes the DFT
of $\Psi_{\nu, q} : \mathcal{J} \rightarrow \mathbb{C}$. 

Let $\mathcal{V}$ denote the set of all the pairs $(\nu, q)$ in the expansion above.
We define the column vector $w_{\vec{k}, \phi} \in \mathbb{C}^\mathcal{V}$ by
$(w_{\vec{k},\phi})_{\nu,q} = \hat{\Psi}_{\nu,q}(\vec{k}) e^{i \nu \phi}$,
and the column vector $z \in \mathbb{C}^\mathcal{V}$ by $(z)_{\nu, q} = \alpha_{\nu, q}$; 
the latter encodes the parameters that describe the target image.
With this notation, $\hat{F}_\phi(\vec{k})$ can be expressed compactly as 
$\hat{F}_\phi(\vec{k}) = z^\top w_{\vec{k},\phi},$ and it follows that
\begin{multline} \label{niceeq}
\hat{S}_3^z(\vec{k}_1,\vec{k}_2) = \\ \sum_{\nu=0}^{6\nu_\mathrm{max}-1} 
\left( z^\top w_{\vec{k}_1,\phi_\nu} \right) \left( z^\top w_{\vec{k}_2,\phi_\nu} \right) 
\left(z^\top w_{-\vec{k}_1-\vec{k}_2,\phi_\nu} \right),
\end{multline}
where we write $\hat{S}_3^z$ to emphasize the dependence on $z$. 
This expression for $\hat{S}_3^z$ is particularly convenient for
computational purposes. In particular, the gradient of $\hat{S}_3^z(\vec{k}_1,
\vec{k}_2)$ with respect to $z$ is easy to compute, which is important for solving the optimization
problem defined in Section~\ref{optprob}.

\subsection{Leveraging symmetries} \label{binstrat}
Recall that $S_3(\vec{x}_1, \vec{x}_2)$ is a discrete version of $s_3(\vec{x}_1, \vec{x}_2)$, 
which only depends on the three parameters: $|\vec{x}_1|$, $|\vec{x}_2|$ and $\theta(\vec{x}_1,\vec{x}_2)$. 
Let $S_3^*$ be the estimate of $S_3$ from the noisy measurement $M$ via 
\eqref{A2S}, that is, the de-biased and normalized autocorrelation of the measurement
$A_3$~\eqref{eq:A3}, and $\hat{S}_3^*$ denote the DFT of $S_3^*$. Since the
noise is assumed to be i.i.d. Gaussian with mean $0$, its Fourier transform is also i.i.d. Gaussian with 
mean $0$ in Fourier space. Therefore, we can reduce the effect of noise by binning
the entries of $\hat{S}_3^*$ with similar values of $|\vec{k}_1|$, $|\vec{k}_2|$ and 
$\theta(\vec{k}_1,\vec{k}_2)$. Numerically, the binning is done through the mapping 
$B :\mathcal{J} \times \mathcal{J} \rightarrow \mathcal{I}$ by
$$
B(\vec{k}_1,\vec{k}_2) = \left( \left\lfloor b_1|\vec{k}_1|
\right\rfloor, \left\lfloor b_1|\vec{k}_2| \right\rfloor
,\left\lfloor b_2 \theta(\vec{k}_1,\vec{k}_2) \right\rfloor \right), 
$$
where $\mathcal{I} \subset \mathbb{Z}_{\ge 0}^3$ represents the set of the bins, 
and $b_1, b_2 \in \mathbb{R}$ determine the bin sizes. 

\subsection{Optimization problem} \label{optprob}
Suppose that we are given 
an estimate of $\hat{S}_3^*$---the de-biased and normalized third-order autocorrelation of the
measurement. For a fixed model and any coefficient vector $z \in
\mathcal{C}^\mathcal{V}$, we can compute $\hat{S}_3^z$ 
via \eqref{niceeq}. In order to estimate the coefficient vector that is consistent with 
$\hat{S}_3^*$, we define the cost function
\begin{equation} \label{eq:opt}
\hspace{-0.3em} f(z) :=  \sum_{\vec{j} \in \mathcal{I}} \left(
\sum_{(\vec{k}_1,\vec{k}_2): B(\vec{k}_1,\vec{k}_2) = \vec{j}} \hspace{-1.2em}
\hat{S}_3^z (\vec{k}_1,\vec{k}_2 ) - \hat{S}_3^* (\vec{k}_1,\vec{k}_2 ) \right)^2\hspace{-0.6em}.
\end{equation}
Minimizing this cost function is a non-convex (polynomial of degree $6$)
least squares optimization problem, and thus, a priori, there is no reason to
suspect that the global minimum of this problem can be
attained; however, our numerical results indicate that standard gradient-based
methods result in accurate and stable recovery of the parameters $z$.
We note that computing the cost and gradient 
involves $\mathcal{O}(n^4 \nu_\text{max})$ operations in each iteration of the 
optimization.

\section{Numerical results} \label{numerical}
\begin{figure}[t!]
\centering
\includegraphics[width=.18\textwidth]{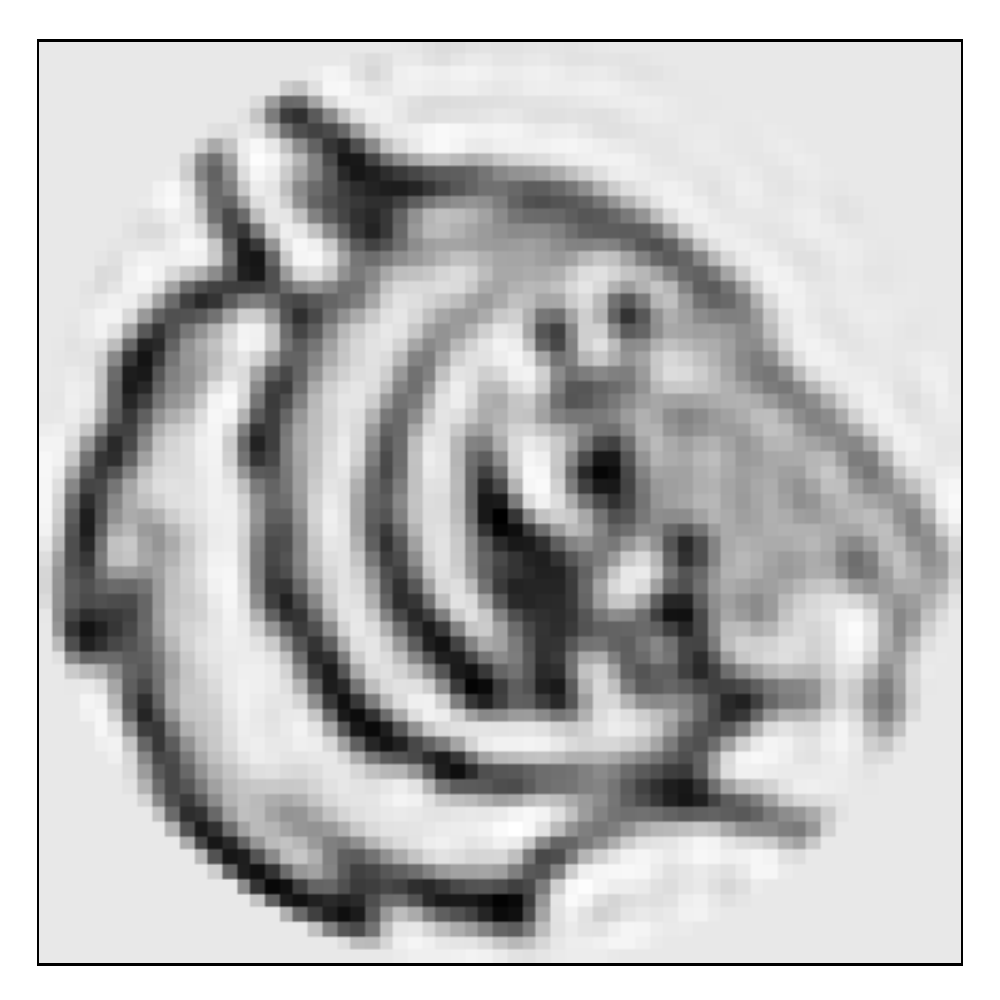} 
\caption{\label{fig1} A $65 \times 65$ image of a tiger expanded in the first
$600$ eigenfunctions, which can be recovered from a noiseless set of
$\hat{S}_3^*$ to high precision.}
\end{figure}

We first consider the problem of recovering a model image from a noiseless set 
of $\hat{S}_3^*$---namely, from the exact rotationally-averaged third-order 
autocorrelation. This reflects the case where the noise level is 
fixed and $m\to\infty$. The model image is generated by expanding a $65 \times
65$ image of a tiger (Figure~\ref{fig1}) into a linear 
combination of the first $600$ Dirichlet Laplacian eigenfunctions, sorted by 
the eigenvalues in ascending order. We denote this model image by $F_0$.
Since $\hat{S}_3^*$ is noiseless in this case, we accelerate the optimization by
only choosing one entry from each bin in $\mathcal{I}$ instead of summing all
the entries to calculate the cost and gradient, which reduces the computational
cost by $\mathcal{O}(n)$. 

Let $F_z$ be the image formed by the coefficients $z$ that were recovered by
minimizing the least squares~\eqref{eq:opt}. Since $\hat{S}_3^*$ 
is invariant under in-plane rotations of the model image, 
we only expect to reconstruct the image $F_0$ up to some arbitrary rotation
$\phi$. As a result, we define the reconstruction error by
$$
\text{error}_\text{recon} := \inf_{\phi \in [0,2\pi)} 
\frac{\|F_0 - F_z^\phi\|_2}{\|F_0\|_2},
$$
where $F_z^\phi$ is the rotation of $F_z$ by an angle $\phi$.
Using the BFGS optimization algorithm, we recovered the image in Figure
\ref{fig1} with 
$\text{error}_\text{recon} = 5\times 10^{-12}$. The optimization took 
$6.5\times10^4$ seconds 
parallelized over 100 CPUs in total. 

\begin{figure}[h!]
\centering
{\footnotesize
$$
\begin{array}{c|c}
\text{error}_{S_3^*} & \text{error}_\text{recon} \\
\hline
1.9\times10^{-4} & 1.6\times10^{-3} \\ 
3.9\times10^{-4} & 3.8\times10^{-3} \\ 
7.7\times10^{-4} & 4.9\times10^{-3} \\  
1.5\times10^{-3} & 1.0\times10^{-2} \\
3.1\times10^{-3} & 2.3\times10^{-2} \\ 
6.2\times10^{-3} & 4.5\times10^{-2} \\ 
1.2\times10^{-2} & 1.1\times10^{-1} \\ 
2.4\times10^{-2} & 1.5\times10^{-1} \\ 
5.0\times10^{-2} & 3.2\times10^{-1} \\
9.9\times10^{-2} & 6.9\times10^{-1} \\ 
\end{array}
\quad 
\raisebox{-.525\height}{\includegraphics[width=.2\textwidth]{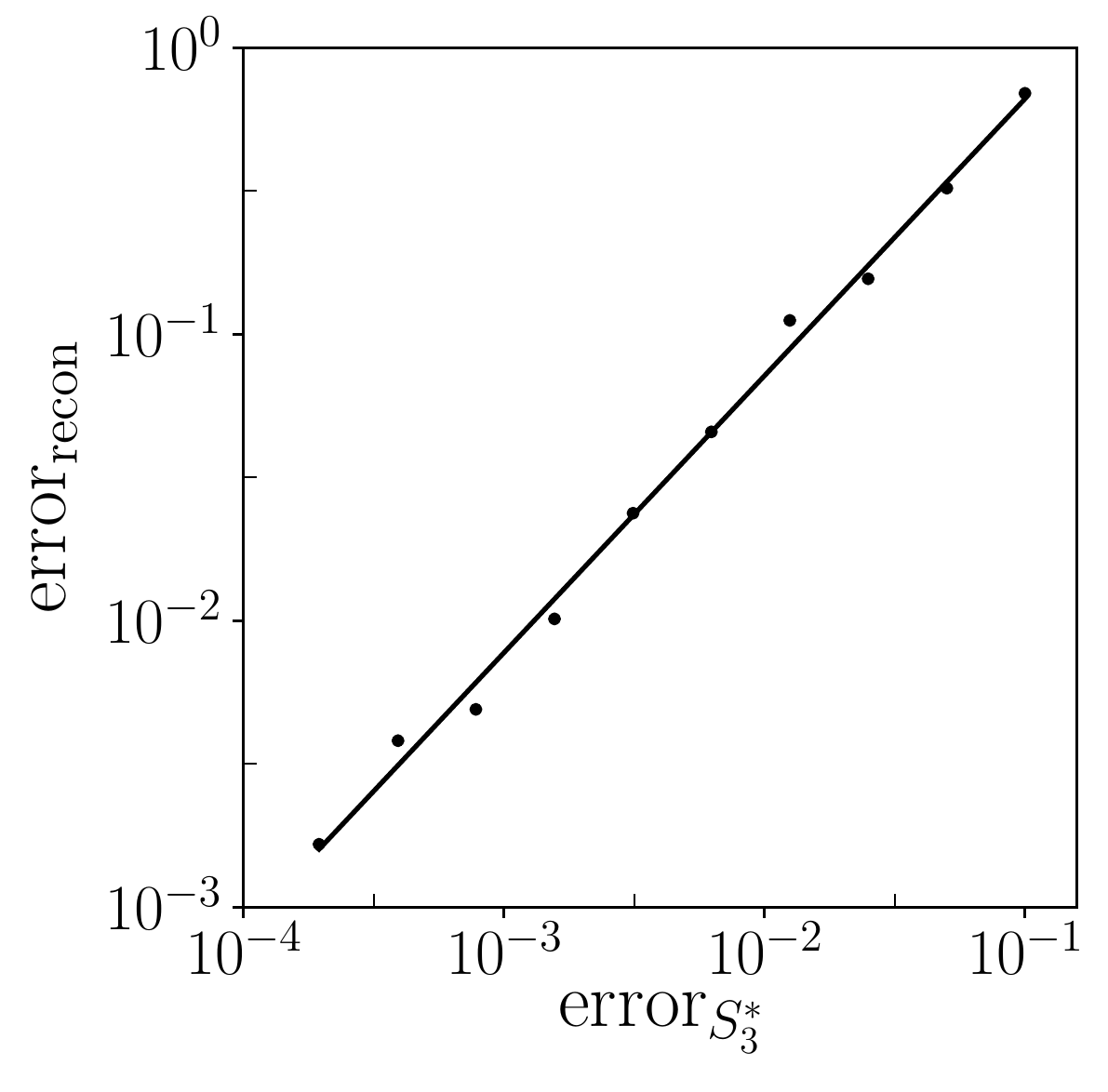}} 
$$
}
\caption{We consider $S_3^*$ with $10$ different levels of Gaussian noise
quantified by $\text{error}_{S^*_3}$. For each level of noise, we run the
optimization five times with random initializations; we report
$\text{error}_\text{recon}$ corresponding to the trial that minimizes the cost
function \eqref{eq:opt}; this procedure can be preformed in practice and makes
the results highly consistent. The slope of the best-fit line is $0.97$.}
\label{fig04}
\end{figure}

In the second experiment, we study the robustness of the reconstruction to noise. 
Limited by the computational resources, we downsample the image of tiger to 
$17 \times 17$ pixels, and expand the image into the first $100$ Dirichlet 
Laplacian eigenfunctions. From the expansion we compute the rotationally-averaged 
third-order autocorrelation $S_3$, which is further contaminated by i.i.d. Gaussian 
noise with mean $0$ and variance $\sigma^2$. A total of 10 different values of 
$\sigma$ are used to model the noisy counterparts $S_3^*$ estimated from 
measurements of different SNR. The relative error of $S_3^*$ is quantified by 
$$
\text{error}_{S_3^*} := \frac{\|S_3 - S_3^*\|_2}{\|S_3\|_2}.
$$
The relative errors of the reconstructed images from the 10 sets of
$\hat{S}_3^*$ are shown in Figure~\ref{fig04}. We observe that the images can be
reliably recovered over a wide range of noise levels. The high correlation
between $\text{error}_{S_3^*}$ and $\text{error}_\text{recon}$ might indicate that
the optimization landscape is benign.

\section{Discussion} \label{sec:discussion}
This work serves as a proof of concept for the feasibility of estimating a
target image from its rotational and translational invariants. We have
demonstrated the reconstruction of a  target image from its noiseless
invariants, and showed that our algorithmic approach is robust to noise.  In
future work, we intend to extend the framework to include the recovery of the
target image from the observation~$M$, to mitigate the separation
condition~\cite{lan2019multi}, and to allow the recovery of multiple images
simultaneously, in a similar fashion to~\cite{bendory2019multi,
boumal2018heterogeneous}. From a theoretical perspective, we wish to complement
the empirical results of this work by proving that indeed a generic image $f$
is determined uniquely by its rotationally-averaged third-order autocorrelation.

\section*{Acknowledgment} 
The authors thank Iris Rukshin for stimulating discussions.  
NFM was supported by NSF DMS-1903015. TYL and AS 
were supported in part by Award Number FA9550-17-1-0291 from 
AFOSR, Simons Foundation Math+X Investigator Award, the Moore 
Foundation Data-Driven Discovery Investigator Award, and NSF 
BIGDATA Award IIS-1837992.


\clearpage
\begin{thebibliography}{00}

\bibitem{aguerrebere2016fundamental}
C.~Aguerrebere, M.~Delbracio, A.~Bartesaghi, and G.~Sapiro, ``Fundamental 
  limits in multi-image alignment,'' \emph{IEEE Transactions on Signal Processing}, 
  vol.~64, no.~21, pp.~5707--5722, 2016.

\bibitem{Bendory2019single}
T.~Bendory, A.~Bartesaghi, and A.~Singer, ``Single-particle
  cryo-electron microscopy: Mathematical theory, computational challenges, and
  opportunities'', \emph{arXiv preprint arXiv:1908.00574}, 2019.

\bibitem{bendory2018toward}
T.~Bendory, N.~Boumal, W.~Leeb, E.~Levin, and A.~Singer, ``Toward single
  particle reconstruction without particle picking: Breaking the detection
  limit,'' \emph{arXiv preprint arXiv:1810.00226}, 2018.

\bibitem{bendory2019multi}
T.~Bendory, N.~Boumal, W.~Leeb, E.~Levin, and A.~Singer, ``Multi-target
  detection with application to cryo-electron microscopy,'' \emph{Inverse
  Problems}, 2019.

\bibitem{bendory2017bispectrum}
T.~Bendory, N.~Boumal, C.~Ma, Z.~Zhao, and A.~Singer, ``Bispectrum inversion
  with application to multireference alignment,'' \emph{IEEE Transactions on
  Signal Processing}, vol.~66, no.~4, pp. 1037--1050, 2017.
  
\bibitem{boumal2018heterogeneous}
N.~Boumal, T.~Bendory, R.~R. Lederman, and A.~Singer, ``Heterogeneous
  multireference alignment: A single pass approach,'' in \emph{2018 52nd Annual
  Conference on Information Sciences and Systems (CISS)}.\hskip 1em plus 0.5em
  minus 0.4em\relax IEEE, 2018, pp. 1--6.

\bibitem{frank2006three}
J.~Frank, \emph{Three-dimensional electron microscopy of macromolecular
  assemblies: visualization of biological molecules in their native
  state}.\hskip 1em plus 0.5em minus 0.4em\relax Oxford University Press, 2006.

\bibitem{grant2018cistem}
T.~Grant, A.~Rohou, and N.~Grigorieff, ``cis{TEM}, user-friendly software for
  single-particle image processing,'' \emph{Elife}, vol.~7, p. e35383, 2018.

\bibitem{henderson1995potential}
R.~Henderson, ``The potential and limitations of neutrons, electrons and
  {X}-rays for atomic resolution microscopy of unstained biological
  molecules,'' \emph{Quarterly reviews of biophysics}, vol.~28, no.~2, pp.
  171--193, 1995.

\bibitem{kakarala2012bispectrum}
R.~Kakarala, ``The bispectrum as a source of phase-sensitive invariants for
  {F}ourier descriptors: a group-theoretic approach,'' \emph{Journal of
  Mathematical Imaging and Vision}, vol.~44, no.~3, pp. 341--353, 2012.

\bibitem{kam1980reconstruction}
Z.~Kam, ``The reconstruction of structure from electron micrographs of randomly
  oriented particles,'' \emph{Journal of Theoretical Biology}, vol.~82, no.~1,
  pp. 15--39, 1980.

\bibitem{Kondor}
R.~Kondor, ``A novel set of rotationally and translationally invariant features
  for images based on the non-commutative bispectrum,'' \emph{arXiv preprint
  arXiv:0701127}, 2007.

\bibitem{lan2019multi}
T.-Y. Lan, T.~Bendory, N.~Boumal, and A.~Singer, ``Multi-target detection with
  an arbitrary spacing distribution,'' \emph{arXiv preprint arXiv:1905.03176}, 2019.

\bibitem{ma2018heterogeneous}
C.~Ma, T.~Bendory, N.~Boumal, F.~Sigworth, and A.~Singer, ``Heterogeneous
  multireference alignment for images with application to {2-D} classification
  in single particle reconstruction,'' \emph{To appear in IEEE Transactions on Image Processing},
  2019.

\bibitem{Mallat}
S.~Mallat, ``Group invariant scattering'', \emph{Communications on Pure
  and Applied Mathematics} vol.~65, no.~10, pp. 1331--1398, 2012.

\bibitem{ozyesil2018synchronization}
O.~\"{O}zye\c{s}il, N.~Sharon, and A.~Singer, ``Synchronization over cartan
  motion groups via contraction'', \emph{SIAM Journal on Applied Algebra and
  Geometry}, vol.~2, no.~2, pp. 207--241, 2018.

\bibitem{punjani2017cryosparc}
A.~Punjani, J.~L. Rubinstein, D.~J. Fleet, and M.~A. Brubaker, ``cryo{SPARC}:
  algorithms for rapid unsupervised cryo-{EM} structure determination,''
  \emph{Nature methods}, vol.~14, no.~3, p. 290, 2017.

\bibitem{sadler1992shift}
B.~M. Sadler and G.~B. Giannakis, ``Shift-and rotation-invariant object
  reconstruction using the bispectrum,'' \emph{JOSA A}, vol.~9, no.~1, pp.
  57--69, 1992.

\bibitem{scheres2012relion}
S.~H. Scheres, ``{RELION}: implementation of a {B}ayesian approach to cryo-{EM}
  structure determination,'' \emph{Journal of structural biology}, vol. 180,
  no.~3, pp. 519--530, 2012.

\bibitem{tang2007eman2}
G.~Tang, L.~Peng, P.~R. Baldwin, D.~S. Mann, W.~Jiang, I.~Rees, and S.~J.
  Ludtke, ``{EMAN2}: an extensible image processing suite for electron
  microscopy,'' \emph{Journal of structural biology}, vol. 157, no.~1, pp.
  38--46, 2007.

\bibitem{tukey1953spectral}
J.~Tukey, ``The spectral representation and transformation properties of the
  higher moments of stationary time series,'' \emph{Reprinted in The Collected
  Works of John W. Tukey}, vol.~1, pp. 165--184, 1953.

\end{thebibliography}
\end{document}